\documentclass[twocolumn, aps, reprint,prl]{revtex4-1}
\usepackage[utf8x]{inputenc}
\usepackage{hyperref}
\usepackage{graphicx}
\usepackage{amsmath}
\usepackage{amsfonts}

\usepackage{url}

\newcommand{\im}{\mathrm{i}}

\newcommand{\unit}[1]{\,\mbox{#1}}
\newcommand{\ket}[1]{\left| #1\right>}
\newcommand{\bra}[1]{\left< #1 \right|}

\begin{document}

\title{All-optical quantum information processing using Rydberg gates}
\author{D. Paredes-Barato$^{\textrm{1\hyperlink{con1}{*}}}$}
\author{C. S. Adams$^{\textrm{1\hyperlink{con2}{*}}}$}
\affiliation{$^{\textrm{1}}$ Department of Physics, Durham University, Rochester Building, South Road, Durham DH1 3LE, United Kingdom}

\date{\today}
 \begin{abstract}
In this work we propose a hybrid scheme to implement a photonic
controlled-z (CZ) gate using photon storage in highly-excited Rydberg
states, which controls the effective photon-photon interaction using resonant
microwave fields. Our scheme decouples the light propagation from the
interaction and exploits
the spatial properties of the dipole blockade phenomenon to realize a
CZ gate with minimal loss and mode distortion. Excluding the
coupling efficiency, fidelities exceeding 95\% are 
achievable and are found to be mainly limited by motional dephasing and
the finite lifetime of the Rydberg levels.   
\end{abstract}


\maketitle


Although optical photons are ideal for quantum communication, their
utility for computation is limited by the lack of strong
photon-photon interactions \cite{Nielsen00, Aaronson11}. However, recently there
has been a substantial progress in this area using Rydberg ensembles 
\cite{Saffman10,Pritchard10, Pritchard13}, where the strong interactions between
highly-excited Rydberg atoms are mapped into strong interactions
between individual optical photons \cite{Dudin12,Peyronel12,Maxwell13,
  Hofmann13, Baur13}. In addition, quantum gate protocols based on
Rydberg atoms have been proposed \cite{Jaksch00} and realised
\cite{Wilk10,Isenhower10} where the information was encoded in the
ground state of the atoms instead of photons. The idea
of exploiting the large dipole-dipole interactions between Rydberg atoms
for photon processing has been analysed theoretically for a variety of
scenarios \cite{Friedler05, Shahmoon11, Gorshkov11, Gorshkov12}. In general
the interaction is dissipative, however dissipation can be reduced at
the cost of interaction strength by detuning off-resonance
\cite{Sevincli11, Parigi12}. An additional problem is the implicit link
between the interaction and propagation, which inevitably leads to a
distortion of the photon wave packet and thereby precludes the
realisation of high fidelity gates, which is one of the requirements for
quantum information processing. In fact it has been argued on
fundamental grounds \cite{Shapiro06, GeaBanacloche10} that this problem
occurs whenever conventional optical non-linearities (such as
cross-phase modulation) are used, and cannot be circumvented.

In this paper, we present a photon gate scheme that decouples light
propagation and interaction, 
allowing the realization of high-fidelity photon-photon gates
with negligible loss or distortion. We use the dark-state
polariton protocol \cite{Fleischhauer00, Gorshkov07} to convert two photonic
qubits (control and target) in the dual rail encoding into collective
excitations with Rydberg character in different positions or \emph{sites} in an
ensemble of cold atoms. We subsequently perform a $2\pi$-rotation on the
target qubit using a microwave field coupled to an
auxiliary state, which by default gives an overall phase shift of
$\pi$-radians to the qubit pair (a Z phase gate). However, the microwave
field also induces resonant dipole-dipole interactions \cite{Maxwell13}
between the target and the control sites that are closest together,
preventing the rotation for one of the four qubit-pair states, and
thereby implementing a CZ phase gate. The excitations are then converted back
to photons and emitted by the ensemble in the phase-matched direction. 

Our scheme relies on the ability to modify the range of the dipole-dipole
interactions between highly-excited Rydberg atoms using a resonant
microwave field \cite{Tanasittikosol11, Bariani12, Maxwell13}. 
By using this field to couple to an auxiliary Rydberg
state, we exploit the spatial independence of the dipole blockade
mechanism  \cite{Sevincli11} to induce a homogeneous phase shift on the
stored photon, and thereby circumvent the local-field limitation of the
optical Kerr effect \cite{Shapiro06, GeaBanacloche10}.  
\begin{figure}[t]
\begin{center}
\includegraphics[width=\columnwidth]{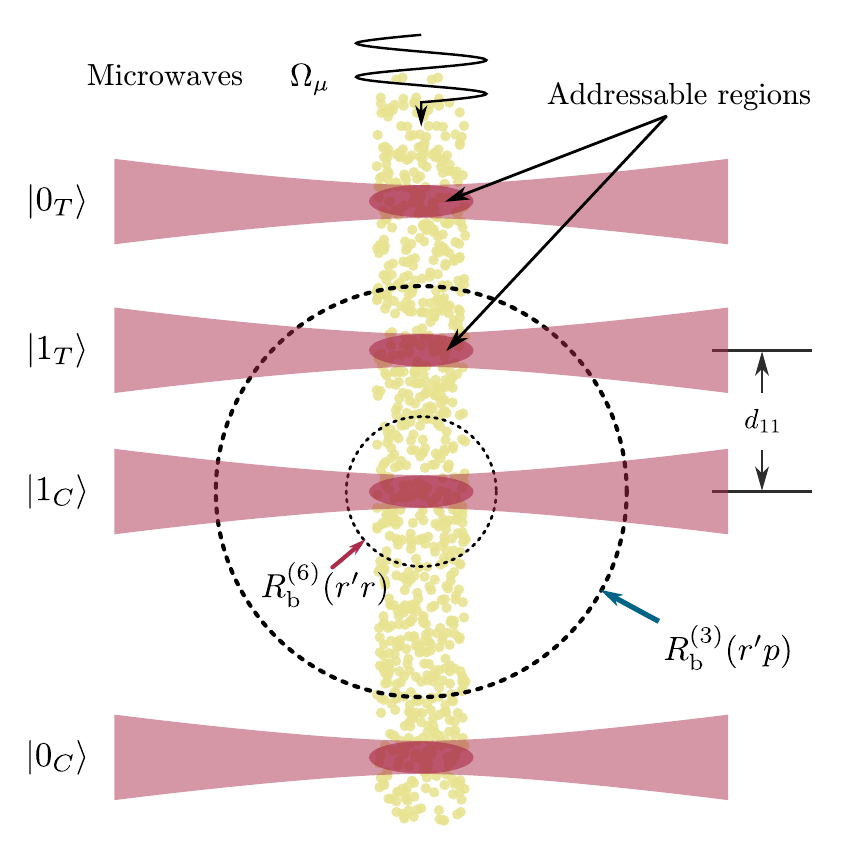}
\end{center}
\caption{Optical layout: Control ($\ket{C}$) and Target ($\ket{T}$)  
  photonic qubits, in dual-rail encoding, are stored as Rydberg
  polaritons (dark red) in a cold atomic ensemble 
  (yellow). The spatial modes corresponding to the qubit states
  $\ket{1_{C}}$ and $\ket{1_{T}}$ are stored in adjacent sites at a
  distance $d_{11}$, and the others arbitrarily further apart. After
  storage we attempt a $2\pi$ rotation on the target qubit using the
  microwave field with Rabi frequency $\Omega_\mu$ and an intermediate
  state. This succeeds except for
  $\ket{1_{C} 1_{T}}$, in which the intermediate state is shifted by
  resonant dipole-dipole interactions, which have a characteristic
  lengthscale $R^{(3)}_{\rm b}$. We need to ensure that the van der
  Waals interactions between stored states (characterized by the
  blockade radius $R^{(6)}_{\rm b}$) are small.} 
\label{fig:layout}
\end{figure}

This Letter is organised as follows: first we outline the storage
procedure, and then we show how, for two photons 
stored in adjacent sites in an atomic cloud, resonant dipole-dipole
interactions can be used to obtain a $\pi$ phase shift to the desired qubit
state. Afterwards, we show that off-resonant, van der Waals interaction
between Rydberg levels in adjacent sites disrupts the ideal process of
the gate, but that its short-range effect can be overcome thanks to the
long range scaling of the resonant interactions. 
Finally, we give an estimate of the gate fidelity for the example of
ultra-cold $^{87}$Rb atoms, where we consider the effects of finite
coupling strengths, extended spatial samples and finite temperature.
   
\begin{figure}[t]
\begin{center}
\includegraphics[width=\columnwidth]{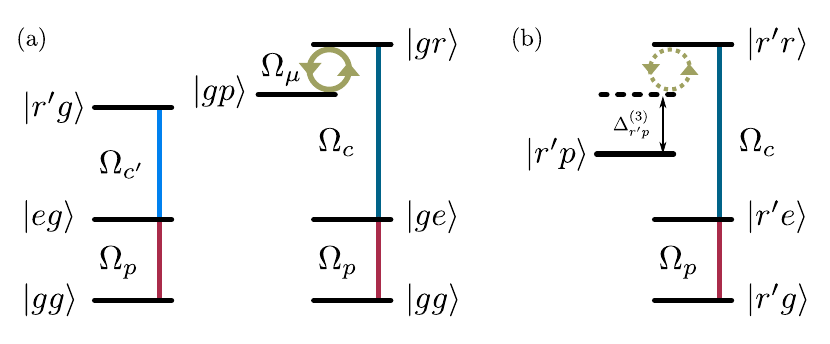}
\end{center}
\caption{Coupled basis of the two inner polariton sites $\ket{1_C
    1_T}$. (a) The control (left) and target (right) photonic qubits are
  stored in the  atom cloud in two different Rydberg states,
  $\ket{r^\prime}$ and $\ket{r}$, respectively. If the control qubit is
  in $\ket{0}$ ($\ket{gX}$ in the coupled basis), we can perform 
  a resonant $2\pi$ rotation on the $\ket{r} \leftrightarrow \ket{p}$
  transition. (b) If the control qubit is in $\ket{1}$ ($\ket{r^\prime
    X}$), it shifts the auxiliary state 
  $\ket{p}$ via a resonant dipole-dipole interactions and the
  microwave $2\pi$-pulse is no longer resonant.  
  This gives rise to a homogeneous, conditional phase shift in the site.
} 
\label{fig:levelsystem}
\end{figure}

The photonic qubit is defined using the dual-rail encoding \cite{Nielsen00},
where the two states of the computational basis in each qubit ($\ket{0}$ and
$\ket{1}$) travel through two spatially separated regions of an atomic cloud.
For a two-qubit gate
we consider four separate spatial 
paths (see \figurename~\ref{fig:layout}).  Similar geometries with only
two sites have already been implemented \cite{Isenhower10, Wilk10}. 
We label the four channels as 
the elements of the set $ \mathbb{B}=\{\ket{1_C}, \ket{0_C},
\ket{1_T},\ket{0_T}\}$, where the subscript represent the (C)ontrol
and (T)arget qubits. We arrange the paths for the $\ket{1}$
(\emph{interacting}) components to be adjacent while the $\ket{0}$
paths are farther apart.
We store the different photonic components in the medium as collective
excitations (also called \emph{dark-state polaritons}) with
Rydberg character using electromagnetically induced transparency (EIT)
in a ladder configuration \cite{Pritchard10,Peyronel12,Maxwell13,
  Hofmann13, Baur13}. To this end, the signal light is resonant with the
closed atomic transition $\ket{g}\leftrightarrow\ket{e}$, and classical
coupling lasers resonant with the transitions $\ket{e} \leftrightarrow
\ket{r}$ or $\ket{e} \leftrightarrow \ket{r^\prime}$  are employed to
store the control and target photons in two different Rydberg states
$\ket{r}$ and $\ket{r^\prime}$ (see \figurename~\ref{fig:levelsystem}). 
We assume that these states are
of the form $\ket{r} = \ket{nS}$ and $\ket{r^\prime} = \ket{n^\prime
  S}$, where $n$, $n^\prime$ are the principal quantum numbers and $S$ denotes the
$L=0$ angular momentum state. Using different Rydberg states for target
and control qubits allows us to perform operations on the individual
qubits using a global microwave field.

The gate works with two photonic qubits, so there is at most one
excitation in each of the sites. The excitation is shared
amongst all the atoms in that site, which maps the state into the
superposition
$\ket{S_j}=\frac{1}{\sqrt{N}} \sum^{N_j}_k |r^j_k\rangle e^{\im \phi_k}~,$
where at each site $j\in \mathbb{B}$ with $N_j$ atoms, the sum spans
all possible singly-excited states $|r^j_k\rangle$ to the Rydberg level
$|r^j \rangle$: $\ket{r}$ in the target qubit,
$|r^\prime \rangle$ in the control. The phase $\phi_k$ depends on the
probe and coupling fields at the position of atom $k$. This process maps
the photonic state $\ket{CT} = \ket{C} \otimes \ket{T}$ into a spin-wave
state involving all of the four spatial channels
$\ket{S_{CT}}=\ket{S_C}\otimes\ket{S_T}$, and can be achieved with
efficiencies per site exceeding 90\% given a sufficiently high optical
depth \cite{Gorshkov12}.

Once we have a mapping of the two-qubit state into the cloud, we make
use of an auxiliary state $\ket{p}$ in the target qubit to perform the
gate operation. A microwave pulse is applied to the system to attempt a
$\int_0^t \Omega_\mu \mathrm{d}t =2\pi$ rotation on the transition
$\ket{r} \leftrightarrow \ket{p}$ in the target qubit, via the
Hamiltonian $H_\mu =\hbar \Omega_\mu
(\ket{r}\bra{p}+\ket{p}\bra{r})$. Since there is
only one excitation at each site, each ensemble behaves like an
effective spin system, coupling the target states $\ket{S_T}$
  and the superposition of singly-excited $\ket{p_k}$ states, $\ket{P_T} =
\frac{1}{\sqrt{N}} \sum^{N_C}_k \ket{p_k} e^{\im \phi_k},$
with the single-atom Rabi frequency $\Omega_\mu$. Since the wavelength
of the microwave field is much greater than the separation between
sites, the coupling to both target sites is the same.

In the absence of other interactions, performing the $2\pi$-rotation
adds a $\pi$-phase shift to the wavefunction, $\ket{CT}\rightarrow
-\ket{CT}$. 
However, if the target state $\ket{p}$ is coupled to the control
Rydberg state $\ket{r^\prime}$ via an electric-dipole interaction at a
distance $d$, dipole-dipole interactions shift the energy of the coupled
state $\ket{r^\prime p}$ by $H_{dd}= \hbar \Delta_{ r^\prime p}
\ket{r^\prime p}\bra{r^\prime p} = (C_3 ( r^\prime p) /d^3
)\ket{r^\prime p}\bra{r^\prime p},$ which can
prevent the rotation, and thus the phase shift, conditional on the
presence of a control excitation in a nearby channel. This operation,
which implements a CZ gate, occurs with
arbitrarily high fidelity if the distance between the adjacent control
and target channels, $d_{11}$, is 
smaller than the characteristic length, 
\begin{equation}
d_{11} < R_{\rm b}^{(3)}(r^\prime p)=\sqrt[3]{C_3(r^\prime p)/\hbar\Omega_\mu}~,
\label{eq:condition2}
\end{equation} 
where $\Omega_\mu$ is the microwave Rabi frequency.

However the discussion outlined above is only valid if there are no
other interactions between sites. If we have two excitations at
a distance $d$ in the medium (one for each qubit), off-resonant, van der
Waals (vdW) interactions between the Rydberg levels $\ket{r}$ and 
$\ket{r^\prime}$ are important, and can hinder the process of the gate
by introducing spatially-dependent detunings to the interacting modes. 
These interactions detune the
doubly-excited state $\ket{r^\prime r}$ by an amount $H_{vdW}= \hbar \Delta_{r^\prime
  r}\ket{r^\prime r} = (C_6(r^\prime r)/d^6 )\ket{r^\prime r} $ by coupling the states
$\ket{r^\prime r} \leftrightarrow \ket{p^\prime p}$, where $\ket{p}$
and $\ket{p^\prime}$ are dipole-coupled to both $\ket{r}$ and
$\ket{r^\prime}$. Here, $C_6(r^\prime r)\propto 1/\delta_f$ is the 
vdW coefficient of an $\ket{n^\prime S,nS}$ pair
state, where $\delta_f$ is the F\"orster energy
defect \cite{Comparat10, Vaillant12}. 

If the energy shift $\hbar \Delta_{r^\prime r}$
is comparable to the energy defect $\delta_f$, dipole-dipole
interactions populate neighbouring states. Also, if this energy defect
is zero (a situation called \emph{F\"orster resonance}), we expect
excitation hopping between $\ket{r^\prime r}$ and $\ket{p^\prime p}$ to
occur spontaneously. Therefore, we need to avoid these situations
choosing an appropriate level system.
 
Even if we have a system without F\"orster resonances, vdW interactions
affect the proper functioning of the gate: during the storage and retrieval
processes and during the rotation in the microwave domain.

If we have an excitation $\ket{r^\prime}$ in one site, the interaction
shift between sites prevents an excitation to $\ket{r}$ within a certain region characterized by the blockade lengthscale, $R^{(6)}_{\rm b}(r^\prime
r)= ({C_6(r^\prime r)/\hbar \Omega})^{1/6}$ , where $\Omega$ is the 
(power broadened) linewidth of the EIT transparency window. 
We minimize these inter-site interactions by ensuring that the distance
$d$ between any two spatial channels satisfies the inequality 
\begin{equation}
d > R^{(6)}_{\rm b}(rr^\prime)~.
\label{eq:condition1}
\end{equation} 
This condition ensures that the interaction
during the storage and retrieval stages of the gate is negligible, thus
avoiding distortion of the spatial modes of the qubits.

Also, the vdW interactions between $\ket{r}$ and
$\ket{r^\prime}$ are present even during the microwave rotation, when
the coupling laser is off. This space-dependent energy shift would cause
a dephasing to the interacting component $\ket{11}$ that would be
proportional to the time $\tau_{2\pi} = 2\pi/\Omega_\mu$ taken to
perform the $2\pi$ rotation, and to $\Delta_{r^\prime
  r}$. But it decreases rapidly with the 
distance between interacting sites. Therefore we need 
achieved for 
\begin{equation}
d_{11} > R_\mu = \left({C_6(r^\prime r)}/{\hbar\Omega_\mu}\right)^{1/6}~.
\label{eq:condition3}
\end{equation}

If we condition the interaction between $\ket{1_C}$ and $\ket{1_T}$
by \eqref{eq:condition2}, and make sure that the effect of vdW interactions are
negligible during the storage/retrieval \eqref{eq:condition1} and the
microwave rotation \eqref{eq:condition3}, the photonic component
$\ket{11}$ picks up a homogeneous $\pi$-phase with respect to
$\ket{00}$, $\ket{10}$ and $\ket{01}$. Then, the overall change in the system
corresponds to that of a CZ-gate \cite{Nielsen00}.

\begin{figure}[t]
\begin{center}
\includegraphics{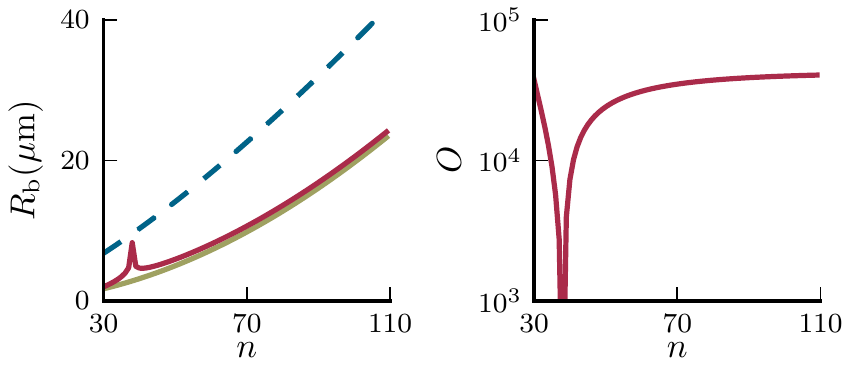}
\end{center}
\caption{(Left) The characteristic
  lengthscales $R_{\rm b}$ as a 
  function of principal quantum number, $n$, for different pair states
  in $^{87}$Rb.  In solid red, the blockade radii for vdW interactions
  $R_{\rm b}^{(6)} (nS_{1/2},(n+1)S_{1/2})$. In yellow, standard vdW
  blockade radii for same-level pair state with coefficient
  $C_6(nS_{1/2},nS_{1/2})$ are shown for reference.  
  In dashed blue, the long-range resonant interactions lengthscale
  $R_{\rm b}^{(3)}(nS_{1/2},nP_{1/2})$ for the $M=0$ pair state.
  All radii  are calculated for a coupling of $1\unit{MHz}$. 
  Note the F\"orster resonance for the coupling $38s39s \leftrightarrow
  38p_{3/2}38p_{3/2}$ \cite{Gallagher09}. (Right) The figure of merit
  $O$ [see \eqref{eq:figureMerit}] for $\ket{nS}$, $\ket{(n+1)S}$ and
  $\ket{nP}$ as a function of the principal quantum number $n$.} 
\label{fig:radii}
\end{figure}

Conditions \eqref{eq:condition2}, \eqref{eq:condition1}, and
\eqref{eq:condition3} suggest using $R_{\rm
  b}^{(3)}/\max(R_{\rm b}^{(6)},R_{\mu}) =
\left( {C_6} \min(\Omega^2,\Omega_\mu^2)/{C_3^2
    \Omega_\mu}\right)^{1/6}$ as the figure of merit, but in
reality, both $R_{\rm b}^{(3,6)}$ and $R_\mu$ are bounded by the
shortest lifetime $\tau=1/\Gamma$ involved in the system. Therefore, we can
optimize the CZ gate operation by choosing a system that maximizes the 
dimensionless figure of merit, 
\begin{equation}
O = \frac{C_3(r^\prime p)^2 }{C_6(r^\prime r) \hbar\Gamma}~.
\label{eq:figureMerit}
\end{equation}
Note that this figure of merit does not depend on any
experimental parameters, and just depends on physical properties of the
atomic species used and the level system chosen. 
Both $R_{\rm b}^{(3,6)}$ and $O$ for $^{87}$Rb are shown as a function of principal
quantum number in \figurename~\ref{fig:radii}.

To better understand the possible implementation of the phase
gate including real-world sources of decoherence, we
estimate the fidelity using a simplified optical Bloch-equation
approach. Our aim is not to provide a full many-body simulation of
the gate protocol, but rather to estimate the errors in the case of a
physical realisation using a cloud of cold $^{87}$Rb
atoms. We shall note that we do not fully simulate storage and retrieval
processes; instead, we use a one-photon transition to the Rydberg states
to this effect (more details can be found in the Supplemental Information).

We choose $\ket{r^\prime}= \ket{nS_{1/2}}$,
$\ket{r}=\ket{(n+1)S_{1/2}}$ and $\ket{p}=\ket{nP_{1/2}}$ to
maximize the ratio $R_{\rm b}^{(3)}(r^\prime p)/R_{\rm
b}^{(6)}(r^\prime r)$ and avoid F\"orster resonances in the region of
interest. For example, for $n=70$, we obtain $R_{\rm  b}^{(6)}\sim
7\,\mu \mbox{m}$ and $R_{\rm b}^{(3)}\sim20\,\mu \mbox{m}$ by coupling
to the $M=0$ state, i.e., the characteristic length of the resonant
microwave transition is around 3 times larger than the optical blockade. 

\begin{figure}[t]
\begin{center}
\includegraphics[width=\linewidth]{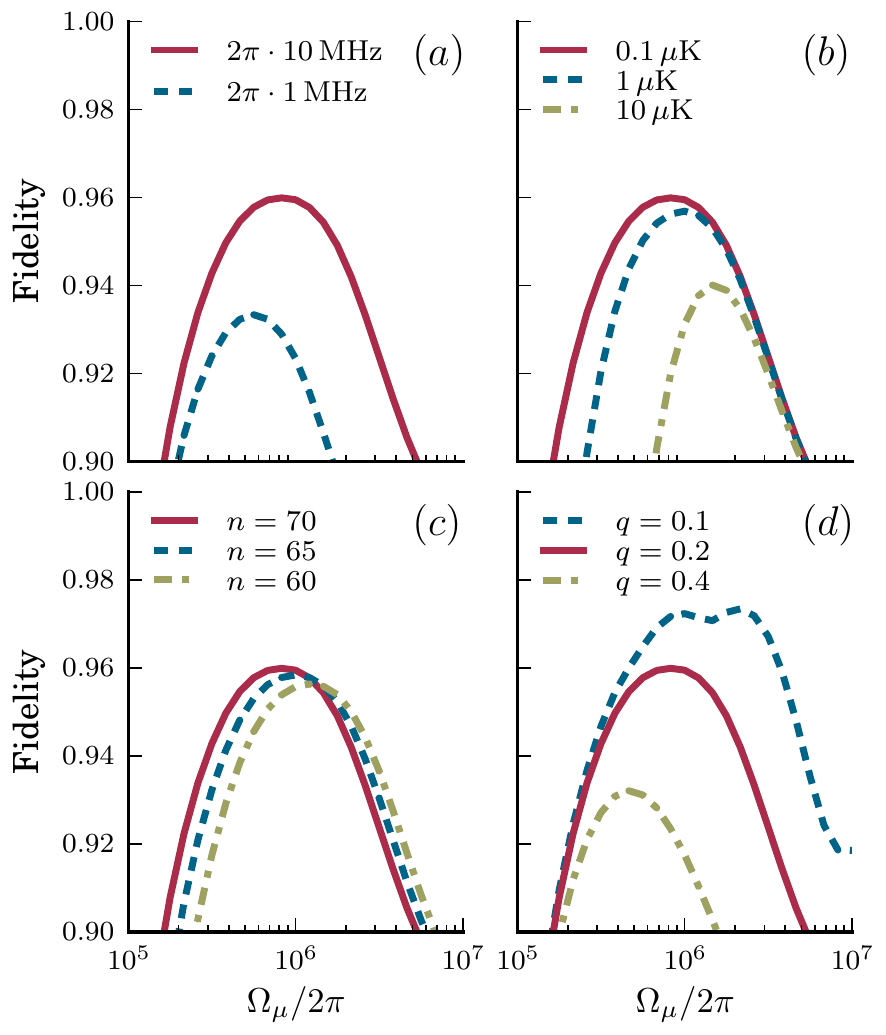}
\end{center}
\caption{Estimation of the fidelity of the gate protocol, including
  the effects of finite lifetimes of the Rydberg states and motional
  dephasing, as a function of the microwave Rabi frequency
  $\Omega_\mu$. The red, continuous line shows the case where $n=70$,
  $\Omega_c = 2\pi \cdot 10 \, \mbox{MHz}$ and $q=0.2$ (see text for
  details), for a temperature of $T=0.1\, \mu\mbox{K}$. Each plate shows
  the changes in the fidelity by varying one parameter.(a) Coupling Rabi
  frequency $\Omega_c$. (b) Temperature $T$. (c) Principal quantum
  number $n$. (d) Different waist to blockade ratios $q = w_0/R_{\rm
    b}^{(6)}$ used in the Gaussian averaging.} 
\label{fig:FinalProbability}
\end{figure}

In \figurename~\ref{fig:FinalProbability} the results of this
exploration are shown, where we have calculated the fidelity $F_0$ for the
initial state $\ket{\psi} =(\ket{00}+\ket{01}+\ket{10}+\ket{11}) /2$ in
the double-qubit basis to become $\ket{\psi^\prime} =
(\ket{00}+\ket{01}+\ket{10}-\ket{11})/2$ after the  
action of the gate. 
We initially obtain $F_0$ as a function of distance, keeping the rest of
the parameters constant. To account for the finite width of the
sites, we convolve $F_0$ with a Gaussian of width $w = \sqrt{2} q
R_{\rm b}^{(6)}(r^\prime r)$, where $q = w_0/R_{\rm  b}^{(6)}(r^\prime
r)$ is the ratio between the probe waist $w_0$ at each 
site and $R_{\rm b}^{(6)}(r^\prime r)$. The $\sqrt{2}$ factor states
that the interaction takes place between two sites. Finally, since the
stored excitations are spin-waves, and these can suffer from motional
dephasing, we multiply the fidelity by a motional dephasing coefficient
$\eta_m \propto \exp \left[-(t^2/\tau^2)/(1+t^2/\xi^2)\right]$ (taken
from \cite{Jenkins12}), where atoms at a temperature $T$ and average
speed $v=\sqrt{k_B T/m}$ ($m$ is the atomic mass) can exit the site with
mode diameter $w_0$ in a time $\xi = w_0/v$, or can move across the
stored spin-wave with wavelength $\Lambda$ in a time $\tau =
\Lambda/2\pi v$. With these factors taken into account, we obtain
processing fidelities over $95\%$ over a broad range of experimental
parameters (see 
\figurename~\ref{fig:FinalProbability}).   

Inspecting \figurename~\ref{fig:FinalProbability} we note the
following general remarks: ensuring a strong coupling $\Omega_c$ is
key, as it allows the two interacting sites to be stored close together, and
profit from a higher resonant dipole shift. Increasing the principal
quantum number increases the fidelity, although we expect a weak scaling
with $n$, as seen inspecting $O$ in
\figurename~\ref{fig:FinalProbability}. This happens because we can
drive transitions in the microwave domain with a weaker $\Omega_\mu$ due
to the favourable scaling of the lifetimes. However, this moves the peak
of the fidelity towards lower driving frequencies making the gate
operation slower, which puts this parameter in competition with motional
dephasing. Finally,  the smaller the waist of the sites, the higher the
fidelity, but this is limited by the diffraction limit; also, when the
sites are very small, achieving a high OD is challenging, and motional
dephasing becomes a problem. 

In addition to the limitations outlined above, a significant source of
inefficiency is likely to arise from the mapping between the light field
and the stored polaritons \cite{Gorshkov12}. However, by making the
cloud sufficiently dense ($N \sim 10^{14}\unit{cm}^{-3}$), it is
possible to obtain optical depths $\mbox{OD}\sim 1000$ that would give
an efficiency per-channel of $\eta_c \approx 0.9$ and an overall \cite{Nunn08}
efficiency $\eta_C^2 \approx 81\%$, although denser
samples might show more dephasing.  This coupling
efficiency can be further increased by using photonic waveguides or by
optimizing the temporal shape of the probe pulse
\cite{Gorshkov12}. 
These numbers compares favourably with previous implementations using
linear optics \cite{Ralph02,OBrien03,Pooley12}, which have a 1/9
efficiency before post-selection, and experimentally achieves $\eta^2
\sim 85\%$ after post-selection.
The process of storage and retrieval of
polaritons with Rydberg content is still not fully understood, and
further optimisations may be possible.

One can imagine using this scheme in combination with integrated chip atom
trapping and waveguides \cite{Kohnen11} to join several quantum gates,
both sequentially and in parallel. Using existing waveguide technology,
one could also implement single qubit operations
\cite{OBrien09,Crespi11} in the same chip, which brings us closer
towards a fully integrated quantum processor. Also, the proposed geometry
and processing method could be extended to implement a photon switch and
other operations.\\ 

In conclusion, we have shown that it is feasible to realize a
quasi deterministic, high-fidelity universal quantum gate for
photons. We circumvent the restrictions of conventional optical
non-linearities by using the non-local dipole blockade effect 
and by separating the propagation and interaction
phases of the gate. We exploit microwave fields to switch between short
range van der Waals interactions and longer range resonant dipole-dipole
interactions, which allows us to achieve a conditional phase shift on
the stored target photon.  Fidelities in 
excess of 95\% are predicted for currently achievable experimental
conditions, and overall efficiencies exceeding 75\% are theoretically
possible. 
Deterministic photon processing will facilitate a wide
range of efficient quantum information protocols.\\

We acknowledge financial support from Durham University, EPSRC and the EU
Marie Curie ITN COHERENCE Network. 
We thank H. Busche, D. Maxwell, D.J. Szwer, M. P. A. Jones,
S. A. Gardiner, N. Henkel and
T. Gallagher for fruitful discussions, and C. L. Vaillant for assistance
with the dipole-dipole interactions code.
\\
\\
\hypertarget{con1}{*}david.paredes@durham.ac.uk
\hypertarget{con2}{*}c.s.adams@durham.ac.uk

\bibliography{phase-gate.bib}

\end{document}